\documentclass{ws-procs9x6}
\usepackage{graphicx}

\def\gsimeq{\mathrel{\hbox{\rlap{\hbox{\lower4pt\hbox{$\sim$}}}\hbox{$>$}}}}
\def\lsimeq{\mathrel{\hbox{\rlap{\hbox{\lower4pt\hbox{$\sim$}}}\hbox{$<$}}}}

\def\chandra{{\it Chandra}}
\newcommand{\lognlogs}{Log~$N$--Log~$S$}
\newcommand{\Hii}{{H{\sc ii} }}
\newcommand{\ergscmq}{erg s$^{-1}$ cm$^{-2}$}
\newcommand{\e}[1]{\cdot 10^{#1}}

\begin{document}

\title{X-ray number counts of star forming galaxies}

\author{PIERO RANALLI, ANDREA COMASTRI \& GIANCARLO SETTI}
\address{Universit\`a degli Studi di Bologna \& INAF, ranalli@bo.astro.it}




\maketitle



The catalogues of optical identifications of X-ray sources in the
\chandra\ Deep Fields surveys (Barger et al.\ 2003, AJ 126, 632) allow
to compute the X-ray \lognlogs\ of normal galaxies in the
flux range $3\e{-17}$--$3\e{-15}$ \ergscmq.  In Fig.~1 we show the
number counts determined with different selection criteria. We find
that the \lognlogs\ is yet undetermined within a factor of $\sim3$,
the reason for this is still under investigation.

The X-ray \lognlogs\ of galaxies at fainter fluxes can be constrained
making use of the linear correlations between X-ray (0.5--10 keV), Far
Infrared (FIR) and radio (1.4 GHz) luminosities discussed by
Ranalli, Comastri \& Setti (2003, A\&A 399, 39).  The
radio counts for the sub-mJy population of faint, 
$z\sim 1$  galaxies were converted into an X-ray
\lognlogs; it is shown in Fig.~1 together with the total observed
X-ray counts.

The FIR and radio luminosity functions (LFs) for normal galaxies may
also be converted to X-ray LFs. The LFs from different surveys
(IRAS/PSC$z$, Takeuchi et al.\ 2003, ApJ 587, L89; ISO/ELAIS, Serjeant
et al.\ 2004, MNRAS submitted; radio 1.4GHz, Machalski \& Godlowski
2000, A\&A 360, 463) were converted to the X-rays and integrated.  We
show the integrated counts in Fig.~2.

\begin{figure}[!t]
   \centering
\includegraphics[height=.3\textheight,width=.8\textwidth]{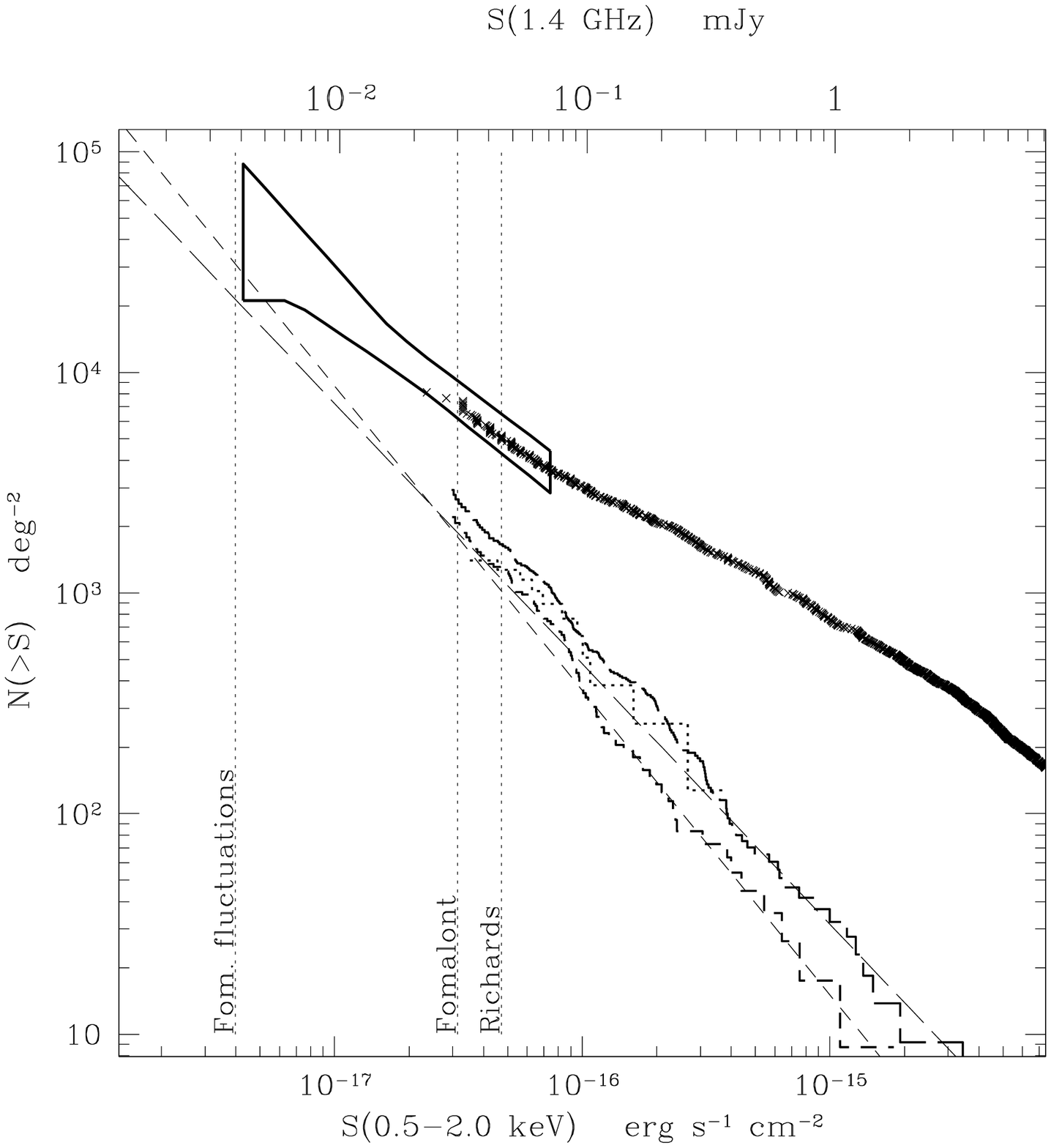}
   \caption{ X-ray number counts from normal galaxies. Thick curve
     and horn-shaped symbol: total X-ray number counts and results
     from fluctuation analysis, respectively.  Vertical dotted lines:
     limiting sensitivities for the radio surveys.  The histograms
     show the observed \lognlogs\ from normal galaxies. Short-dashed
     (lower) histogram: sources with Log (X-ray/optical flux ratio)
     $<-1$. Dotted (middle) histogram: X-ray sources with a radio
     detection and an \Hii region-like optical spectrum (Bauer et al.
     2002). Long-dashed (upper) histogram: sources from the Bayesian
     sample of Norman et al. (2004).  Long-dashed and short-dashed
     lines: radio counts for the sub-mJy population (Fomalont et al.\ 
     1991, AJ 102, 1258 and Richards 2000, ApJ 533, 611,
     respectively) converted to the X-rays.  }
    \label{hard}\bigskip
\includegraphics[height=.3\textheight,width=.8\textwidth]{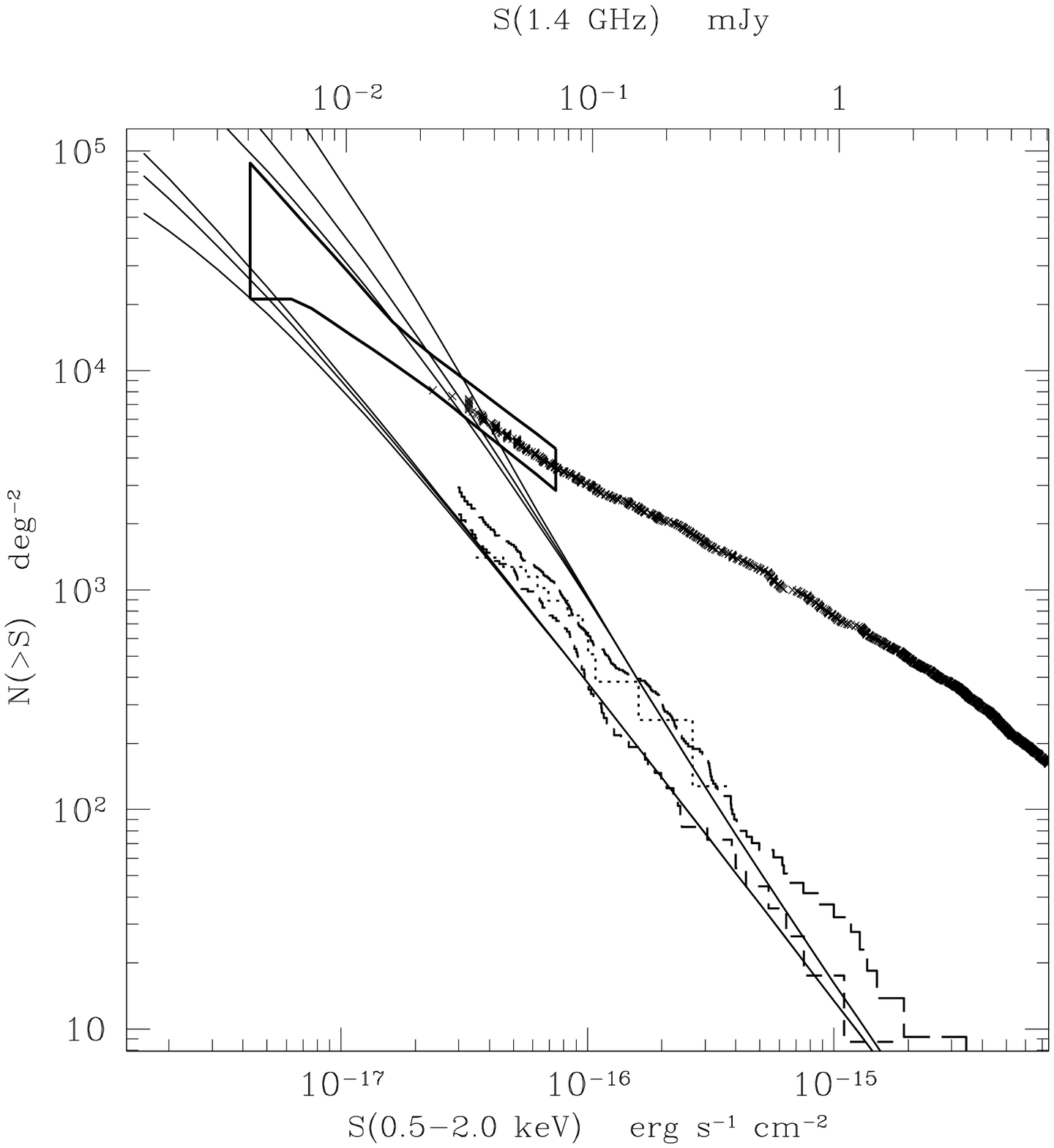}
   \caption{  Solid curves: X-ray counts derived from
     integration of the ELAIS 90$\mu$ and IRAS 60$\mu$ luminosity
     functions (lower triplet: Serjeant et al.\ 2004; upper triplet:
     Saunders et al.\ 1990, MNRAS 242, 318), converted to the X-rays.
     The main difference between the two LFs is the evolution
     ($(1+z)^3$ for Serjeant's and $(1+z)^{6.7}$ for Saunders'). The
     lower and upper curves of each triplet show the integration to
     $z_{\rm max}=1.1$ and 2 respectively, with evolution $(1+z)^3$.
     Middle curve: $z_{\rm max}=2$ but evolution stopped at $z=1$.
     Other symbols as in Fig.~1.  Similar results might be obtained
     from integration of the radio (1.4GHz) luminosity function of
     star forming galaxies (Machalski \& Godlowski 2000). The stronger
     evolution found by Saunders et al.\ is not consistent with both
     the observed counts and the limits from fluctuation analysis.  }
 \end{figure}

The most important results are:
\begin{itemize}
\item the X-ray \lognlogs\ may be expressed as $N(>S)\propto
S^{\sim(-1.3)}$ in the flux interval $3\e{-18}$--$3\e{-15}$ \ergscmq; 
\item the fraction of bona fide star forming galaxies among X-ray
  sources in current deep surveys (at a flux limit of $5\e{-17}$
  \ergscmq) is about 20\%;
\item star forming galaxies are expected to become the dominant
population among X-ray sources at fluxes fainter than 1--2$\e{-17}$
\ergscmq.
\end{itemize}

\end{document}